\documentclass[% twocolumn,
aps,nofootinbib,
showpacs,showkeys,preprint
tightenlines,preprintnumbers,] {revtex4}

\usepackage{epsf,epsfig,subfigure,graphicx,amsmath,amssymb %,mathtools
}
\usepackage{color}
\newcommand{\dis}[1]{\begin{equation}\begin{split}#1\end{split}\end{equation}}
\newcommand{\ie}{{\it i.e.~}}

 \newcommand{\dell}{\delta_{\rm PMNS}}
\newcommand{\delq}{\delta_{\rm CKM}}

\newcommand{\delL}{\delta_{\rm L}}
\newcommand{\delx}{\delta_{\rm X}}
\newcommand{\eL}{\epsilon_{\rm L}}

\def\sw0{{$\sin^2\theta_W^0$}}

\newcommand{\Z}{{\bf Z}}

\def\sm{{SU(2)$_L\times$U(1)$_Y$}} 
\def\E6{{\rm E_6}}

\def\EE8{{\rm E_8\times E_8'}}

\begin{document}

\draft

\title{Leptogenesis with high-scale electroweak symmetry breaking and an
    extended Higgs sector} 

\author{Laura Covi}
\address
{Institut f$\ddot{\rm u}$r Theoretische Physik, Friedrich-Hund-Platz 1, D-37077 G$\ddot{o}$ttingen, Germany
}
\address{CERN, CH�1211 Geneva 23, Switzerland}
\author{Jihn E.  Kim}
\address
{Department of Physics, Kyung Hee University, 26 Gyungheedaero, Dongdaemun-Gu, Seoul 02447, Republic of Korea, and\\
Center for Axion and Precision Physics Research (IBS),
  291 Daehakro, Yuseong-Gu, Daejeon 34141, Republic of Korea
}
\author{Bumseok Kyae}
\address
{Department of Physics, Pusan National University, 2 Busandaehakro-63-Gil, Geumjeong-Gu, Busan 46241, Republic of Korea}

\author{Soonkeon Nam}
\address
{Department of Physics, Kyung Hee University, 26 Gyungheedaero, Dongdaemun-Gu, Seoul 02447, Republic of Korea}

\begin{abstract} 
We propose a new scenario for baryogenesis through leptogenesis, where the CP phase relevant for leptogenesis is connected directly to the PMNS phase(s) in the light neutrino mixing matrix. The scenario is realized in case only one CP phase appears in the full theory, originating from the complex vacuum expextation value of a standard model singlet field. In order to realize this scheme, the electroweak symmetry is required to be broken during the leptogenesis era and a new loop diagram with an intermediate $W$ boson exchange including the low energy neutrino mixing matrix should play the dominant contribution to the CP violation for leptogenesis. In this letter, we discuss the new basic mechanism, which we call type-II leptogenesis, and give an estimate for maximally reachable baryon asymmetry depending on the PMNS phases.
 
\end{abstract}
\maketitle

%%%%%%%%%%%%%%%%%%%%%%%%%%%%%%%%%%%%%%%%%%%%%%%%%%%%
%%%%%%%%%%%%%%%%%%%%%%%%%%%%%%%%%%%%%%%%%%%%%%%%%%%%

\section{Introduction}\label{sec:Introduction}

The origin of the baryon asymmetry of the Universe (BAU) has been a longstanding theoretical issue~\cite{Ibarra02}. 
Among Sakharov's three conditions \cite{Sakharov67} for successful generation of an asymmetry from a symmetric initial
state, the first, i.e. baryon number violation, and the second, i.e. C and CP violation, rely most strongly
on model building beyond the Standard Model (SM) of particle physics.
Along this line, there already exist plenty of theoretical models to generate the 
BAU~\cite{BAgut,EWgen,AD85,Fukugita86,QuaG}, with different
ways to depart from thermal equilibrium, e.g. from heavy particle decay outside equilibrium to first-order phase-transitions or to the dynamical Affleck-Dine (AD) mechanism  \cite{AD85}.

In this letter, we would like to follow an alternative route, realizing instead the leptogenesis mechanism
within a phase with broken electro-weak symmetry at high temperature and relying mostly on
SM physics in the neutrino sector to achieve the necessary CP-violation. 
The only ingredients beyond the SM that we need is the presence of various species of
SM singlets, with the quantum numbers of right-handed(RH) neutrinos, different Higgs doublets, in
order to allow for a non-vanishing contribution to the CP asymmetry from a $W$-boson loop and  to keep the electroweak symmetry broken.

Indeed, at the level of the SM of particle physics, CP violation is related to the charged-current interaction and determined by two CP phases, one in the quark sector, the 
Cabibbo-Kobayashi-Maskawa~(CKM) phase $\delq$ \cite{Cabibbo63,KM73} and the other the 
Pontecorvo-Maki-Nakagawa-Sakada~(PMNS) phase $\dell$  \cite{PMNS}, while two more Majorana phases  
are appearing in the leptonic  sector. 
In a family unified grand unified theory (GUT), these two phases can be related if only one single
complex vacuum expectation value(VEV) appears  in the full theory \cite{KimEPJC15}. 
From the early time on, it has been an 
interesting issue to investigate a possibility of relating the baryon asymmetry with the SM phase(s) $\delq$ or/and $\dell$. 
 
The first obvious possibility is to exploit $\delq$ for the BAU, but it has been known  long time  that such phase, appearing always with small mixing angles, is not enough for the baryon number generation
in GUT baryogenesis \cite{BSW79}. Even in other scenarios, relying on the quark sector, like the AD mechanism
 \cite{AD85} through baryon number carrying scalars or the electroweak baryogenesis, additional CP violating
phases are needed to provide a large enough baryon asymmetry  \cite{Gavela94}.
Therefore, it is difficult to explain the BAU just considering the CP violation in the quark sector.

A more promising road is given by baryogenesis through leptogenesis \cite{Fukugita86}, since the phases 
in the leptonic sector are unconstrained and the mixings large. This mechanism relies on the fact that sphaleron processes are effective before/during the electroweak phase transition and violate both baryon~($B$) and lepton~($L$) numbers, but conserve baryon minus lepton number~($B-L$). Therefore, baryogenesis or leptogenesis above 
the electroweak scale must generate a non-vanishing $B-L$ number, that is then translated into a baryon asymmetry
before or at the electroweak transition.

In this paper we consider a leptogenesis scenario, which allows us to relate the CP-violation during
leptogenesis to the phase $\dell$ in the light neutrino mixing matrix. In order to be able to 
have a well-defined neutrino mixing matrix when the lepton asymmetry is cosmologically created, we
require that the SM gauge group  \sm~ remains broken during the 
leptogenesis epoch. In fact, the Brout-Englert-Higgs(BEH) mechanism for \sm~ breaking at high temperature 
is possible for  some regions in the parameter space of BEH bosons $h_u$ and $h_d$ \cite{Senj79}. 
\vspace{1cm}

 %%%%%
\section{A new type of leptogenesis}\label{sec:Type2}

In the leptogenesis scheme with one or two BEH doublets, the lepton asymmetry arises 
from the decay of   the lightest heavy Majorana neutrino $N_1$ 
producing light leptons and antileptons and the Higgs particle by the decay
\begin{equation}
N_1 \rightarrow \ell_i + h_u \; , \; \bar{\ell}_i + h_u^*\; , 
\label{lepto-Type-I}
\end{equation}
where $\ell_i (\bar{\ell}_i)$ is the i-th lepton (antilepton) doublet and $h_u$ is the up-type $Y = 1/2$ BEH doublet. 
We follow here the supersymmetric notation, but the mechanism can work also without supersymmetry.
In this case therefore the same mother particle $N$ has two decaying channels with different lepton number and therefore the model satisfies the Nanopoulos-Weinberg theorem  \cite{NanoWein79,NanoWein14}.
In classical leptogenesis, the CP violation in the decay arises from the interference of the tree-level with the
one-loop diagrams involving the heavier RH neutrinos $N_j$, $j = 2,3 $  
(for the case of  three generations) and the  CP violation arises in general from the complex Yukawa 
couplings and has in general no direct relation to the low-energy CP-phases~\cite{Branco:2002xf}, apart in case of particular 
textures~\cite{Abada:2006ea} or CP conservation in the heavy RH neutrino sector \cite{Abada:2006ea, Pascoli:2006ci}.

In this letter we would like to extend the model in order to have a large contribution to the CP violation from an electroweak loop involving  explicitly the PMNS matrix. In order to do so, we introduce another copy of the
Higgs doublet $ H_u $, heavier than the SM one $h_u$ and with vanishing vacuum expectation value (VEV), 
as well as another generation of RN neutrinos $ {\cal N}_1$. All these particles can mix with $h_u, N_1$ respectively 
and allow for the presence of the diagram (2b) in Fig.~2, where the virtual particles are all SM particles
and one of the vertices include the PMNS matrix directly.
We consider here for simplicity the case where the field $ H_u $ is heavier than the right-handed neutrino,
so that the decay of $ N $ into $ \ell_i + H_u $ is negligible.
 
The CP phase in the PMNS matrix is required to descend down from high energy scale by a complex VEV. To relate different phases, we assume that only one SM singlet field $X$ develops a CP phase $\delx$. Thus, all Yukawa couplings and the other VEVs are real and all CP violation parameters arise from $\delx$.  

While the SM Higgs doublet(s) do not carry lepton number, we define the fields $H_{d}$ and $H_u$ to carry the lepton number $L=+2$ and $-2$, respectively, and  ${\cal N} $ instead to have $L=1$, while $N$ will be 
defined to carry $L=-1$. We can then write for the Higgs doublets and the heavy neutrinos the Yukawa couplings:
 \dis{
 f N_1  h_u \ell_L, \quad\quad  \tilde{f} {\cal N}_1 H_u \ell_L. 
 \label{eq:Higgs} 
 }
 The Yukawa couplings ($\tilde{f}$'s) of the inert Higgs doublets $H_{u,d}$ to the lepton doublets  are distinguished from those (${f}$'s)  of the BEH doublets $h_{u,d}$ and no mixing is allowed at this level due to the different lepton
 number assignments. We have as well other lepton number conserving interactions as
\dis{
 \Delta m_0\,N_1\, {\cal N}_1 + \mu_H^2 H_uH_d+{\rm H.c.},\label{eq:Ldef}
}
where $\Delta m_0$ is real. The first term of (\ref{eq:Ldef}) gives directly a Majorana mass term between 
$ N_1 $ and $ {\cal N}_1$ without a phase because we defined it preserving the lepton number.
 The Dirac mass for the seesaw neutrino mass is via $N_1 h_u\ell_L$ which appears as in the `Type-I leptogenesis' .
 The needed lepton number violating couplings are introduced by the couplings
\dis{
\Delta{\cal L}\ni \mu^{\prime\, 2} h_u^*H_u+m_0'\,N_1 N_1 + m_0^{\prime\prime}\,{\cal N}_1\,{\cal N}_1+\rm h.c. \label{eq:Lviol}
}
which allow for mixing also in the Higgs sector and for the see-saw mechanism.
There are more $L$ violating terms such as $h_dH_u, h_uH_d$ and $h_d^*H_d$, which are not relevant
for leptogenesis. We will assume $\Delta m_0  \gg m_0^\prime, m_0^{\prime\prime}$, \ie
the $L$ conserving mass parameter is much larger than the $L$ violating mass parameters. In this case, 
$(N_1, {\cal N}_1)$ are maximally mixed and we call $N$ the lightest mass eigenstate obtained from the
mixing of these two states. We can then define an effective Yukawa coupling for this lightest RH neutrino as
\begin{equation}
 f^{eff} \; N  h_u \ell_L, \quad\quad \mbox{with} \quad\quad f^{eff} = f \cos\theta_N + \tilde f \sin\theta_N 
 {\mu^{\prime\, 2} \over m^2_H - m^2_h}
 \label{eq:Higgs2} 
 \end{equation}
by considering the large mixing angle $ \theta_N$ between the neutrinos and 
 also the small mixing between $H_u$ and $h_u$.

  %%%%%
\section{With one phase}\label{sec:OnePhase}

%%%%%%%%%%%%%%%%%%%%%%%%%%%%%%%%%%%%%%%%%%%%%%%%%%%%%%%%%%%%
\begin{figure}[!t]
\begin{center}
\includegraphics[width=0.6\linewidth]
{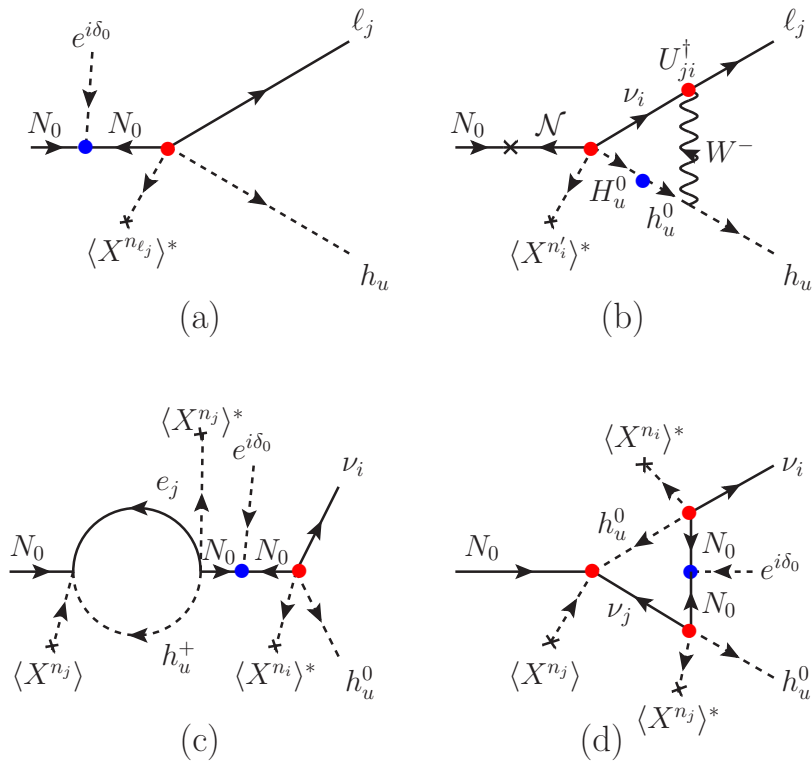}
\end{center}
\caption{The Feynman diagrams interfering in the $N$ decay: (a) the lowest order diagram, (b) the $W$ exchange diagram, (c) the wave function renormalization diagram, and (d)  the heavy neutral lepton exchange diagram.  There exist similar ${\cal N}$-decay diagrams.   In all figures, the final leptons can be both charged leptons and neutrinos. 
 The lepton number violations are inserted with blue bullets and phases are inserted at red bullets. (a) and (b) interfere.  (c) and (d) give a vanishing contribution in $N_0$ domination with one complex VEV.} \label{fig:IntCP}
\end{figure}
%%%%%%%%%%%%%%%%%%%%%%%%%

The process (\ref{lepto-Type-I}) can include the phase $\delx$ by the interference terms with the
diagram with an intermediate $W$-boson. To relate the leptogenesis phase $\delL$ to the SM phase(s), one needs a families-unified GUT toward a calculable theory of the physically measurable phases. In the anti-SU(7) \cite{KimJHEP15}, indeed $\dell$ and $\delq$ are shown to be related \cite{KimEPJC15}.  In this paper, we attempt 
to relate the phases in leptogenesis and $\dell$ \cite{Branco03}. In other words, we attempt to express the lepton asymmetry $\eL$   in terms of $\dell$.  For this,  the $W$-boson loop must dominate over the other
one-loop corrections.

To obtain a calculable theory for phases, we introduce a single Froggatt-Nielsen(FN) field \cite{FN79}  $X$ developing 
a complex VEV, $\langle X\rangle = x\,e^{i\delx}$ \cite{KimPLB11,KimPLB15}. The Yukawa coupling matrix of the 
doublet $h_u$ include powers of $X$ such that some symmetry behind the Yukawa couplings is satisfied.  
 The Yukawa couplings of the three RH 
neutrinos obtain then a complex phase from different powers of the Froggatt-Nielsen field $X$ depending on the
generation. 
To simplify the discussion, let us assume that the heavy Majorana neutrinos have a mass hierarchy and let the lightest heavy Majorana neutrino $N$ dominate in the leptogenesis calculation. 

For the tree $\Delta L \ne 0$ decay mode corresponding to Fig. \ref{fig:IntCP}\,(a), we show the relevant Feynman diagrams interfering in the $N \to \ell_j +h_u$ decay in Fig. \ref{fig:IntCP}\,(a) and (b)   giving rise to a
new contribution to leptogenesis.
In Fig. \ref{fig:IntCP}\,(a), (c), and (d), we also show the relevant diagrams for $N \to \ell_i +h_u$ decay in the  classical leptogenesis scenario, discussed in \cite{Covi96,SarkarU95,Plumacher97}.   In the basis where the $N$s and the charged leptons masses are diagonal, possible phases appear at the vertices with the red bullets in Fig. \ref{fig:IntCP}.  
In models where a single complex VEV appears in all the Yukawa couplings, even if with different powers,
the classical leptogenesis diagrams given in Fig. \ref{fig:IntCP}\,(c) and (d) do not contribute to the CP
asymmetry because the overall phases cancel out with that of Fig. \ref{fig:IntCP}\,(a). Indeed, we see that in the 
diagrams the directions of  $\langle X^{n_j}\rangle$ are opposite, indicating that the phases are equal and opposite.  
We are then left to compute the contribution from the diagram with an intermediate $W$. As we will see
explicitly below, such contribution vanishes in the presence of a single Yukawa coupling, 
but gives a non-vanishing contribution in our model.
 
Let us calculate the effect of the vertex correction via the $W$ boson of Figs. \ref{fig:IntCP}\,(b), in the
simple mass-insertion formalism. A new conclusion will be drawn from this calculation.  
With this set-up, it is a standard procedure to calculate the asymmetry $\eL$, \ie the difference of $N$ decays to the 
$\ell$ and $\bar{\ell}$,
\dis{
\eL^{N}(W)=\frac{\Gamma_{N\to\ell}- \Gamma_{N\to\bar{\ell}}}{\Gamma_{N\to\ell}+ \Gamma_{N\to\bar{\ell}}},
}
where $\ell(\bar{\ell})$ is a (anti-)lepton. We have the following  interference term from Figs. \ref{fig:IntCP}\,(a) and (b),
  \dis{
 \int \frac{d^4k}{(2\pi)^4}\, {\cal M}_{\rm (b)}{\cal M}^\dagger_{\rm (a)}=
i\frac{f_j U_{ji}^\dagger \tilde{f}_i^* g_2^2\,\mu^{\prime\,2}}{
 2 \sqrt{2}(m_{H^0}^2-m_{h^0}^2)}\times 
{\cal I} (P, p_\ell, m_W^2, m_h^2, m_{\nu_i}^2)
\label{eq:integ}
}
where $ {\cal I} $ denotes the loop integral depending on the internal masses and the external
momenta. 

Here we denote with $ \{P,m_0\} $ the four-momentum and mass of $N$, with
$ \{p_{l},m_{l}\} $ the four-momentum and mass of the final state lepton $ l_j$,  
while $m_{\nu_i}, m_{h^{0(+)}}, m_{H^0}$ and  $m_W$ are the masses of the states
$\nu_i, h_u^{0(+)}, H_u^{0}$ and the mass of  the $W$-boson respectively.  
The $Z$ and photon couplings are flavor diagonal and do not contribute. 
Note that the factor ``$\mu^{\prime\,2}/ 
(m_{H^0}^2-m_{h^0}^2)$'' in Eq.~(\ref{eq:integ})
is the mixing angle between $h_u^0$ and $H_u^0$
in the mass eigenbasis when $\mu^{\prime\,2}\ll m_{H^0}^2, m_{h^0}^2$.   
Through such Higgs flavor change at the blue bullet in Fig. 2\,(b),  
the lepton number is violated. 
As in the classical case, the loop integral is UV divergent, but its imaginary part is finite and quite simple
in the limit of vanishing mass for the leptons and $m_W \ll m_0 $. 
Indeed we obtain
\begin{equation}
{\rm Im} [ {\cal I} (m_0^2, m_{h^+}^2, m_W^2, m_{h^0}^2, 0)] \simeq  \frac{m_0^2 - m_{h^0}^2}{8 \pi} \left[ 1 - 
\ln \left( 1 + \frac{m_0^2}{m_W^2} \right) \right],
\end{equation}
where we have used $ 2 P \cdot p_\ell = m_0^2 - m_{h^{+}}^2 $ as set by the kinematical constraints.
This expression is IR divergent for vanishing $m_W$, but in that limit the PMNS matrix $ U_{ij} $ becomes
trivial and the CP violation vanishes automatically.
Indeed considering also the neutrino final states in Figs. \ref{fig:IntCP}\,(a) and (b), in which case the 
PMNS matrix is $U$ instead of $U^\dagger$, 
and the analogous diagram with the $W$-loop attached
to the tree-diagram in Fig. \ref{fig:IntCP}\,(a), we obtain
\dis{
\epsilon_L^{N_0}(W) =
\frac{\alpha_{\rm em}}{2\sqrt{2}\sin^2\theta_W}
\frac{1}{\sum_i | f^{eff}_i |^2}~{\rm Im} \left[ \sum_{i,j}
\,\left[   f^{eff}_j U_{ji}^\dagger (f^{eff}_i)^* 
+   f^{eff}_j U_{ji}  (f^{eff}_i)^*
\right] \right]
\times\Big[1+{\rm ln}\left(\frac{m_W^2}{m_0^2}\right)  
\Big].\label{eq:asymmExp}
}
The asymmetry (\ref{eq:asymmExp}) in the limit of unbroken SU(2), but $m_{h^0,H^0} = m_{h^+,H^+} $, 
  is given by the simple matrix multiplication $f_1^\dagger (U+U^\dagger ) f_2$ where $f_{1,2}$ are column vectors. 
 The imaginary part has the form $\frac12(f_1^\dagger  (U+U^\dagger )   f_2-f_2^\dagger (U+U^\dagger )  f_1)$ 
 which is zero  if $f_1=f_2$ or $(U+U^\dagger )$ is diagonal. This is consistent with the fact that in these diagrams 
 the lepton violation is on the left side of the cut as discussed in \cite{Adhikari:2001yr}.
 Nevertheless in our case thanks to SU(2) breaking, the masses of the particles in the loop are different, so 
 the loop factors are not exactly equal and an imaginary part is present. Indeed expanding for example in the 
 Higgs mass difference $ m_{h^{0/+}} = m_h^2 \mp {1\over 2} \Delta m_h^2 $, we obtain instead
 \begin{equation}
\epsilon_L^{N_0}(W) =
\frac{\alpha_{\rm em}}{ {2}\sqrt{2}\sin^2\theta_W} \frac{\Delta m_h^2 }{m_0^2}
\frac{1}{\sum_i  { | f^{eff}_i |^2}}~{\rm Im} \left[ \sum_{i,j}
\,\left[   f^{eff}_j U_{ji}^\dagger (f^{eff}_i)^*  -  f^{eff}_j U_{ji}  (f^{eff}_i)^*  \right] \right]
\times\Big[1+{\rm ln}\left(\frac{m_W^2}{m_0^2}\right)  
\Big].\label{eq:asymmExp2}
\end{equation}
 If the SU(2)$\times$U(1)$_Y$ is broken, by some choice of BEH boson couplings 
 {\it \`a la} Ref. \cite{Senj79}, the mass splitting is $ \Delta m^2_h \propto v^2(T) $ and
 a substantial CP asymmetry is present at $ T \leq m_0$. Moreover it is then possible to relate 
 the lepton asymmetry directly to $\dell$. 
   
For concreteness, we use the PMNS matrix $U_{ij}$, in the vertex diagram of Fig. \ref{fig:IntCP}\,(b), presented in \cite{KimSeo11,KimEPJC15} together with Majorana phases $\delta_{a,b,c}$,
\dis{
 U&=\left(
\begin{array}{ccc}
c_1 & s_1c_3 & s_1s_3  \\
-c_2s_1 & e^{-i\dell}s_2s_3+c_1c_2c_3 & -e^{-i\dell}s_2c_3+c_1c_2s_3  \\
-e^{i\dell}s_1s_2 & -c_2s_3+c_1s_2c_3 e^{i\dell} & c_2c_3+c_1s_2s_3e^{i\dell} \\
\end{array}\right)_{\rm KS}
\left(\begin{array}{ccc} e^{i\delta_{a}}&0&0\\
0&e^{i\delta_{b}}&0\\ 0&0&~e^{i\delta_{c}} 
\end{array}\right)_{\rm Maj},\label{eq:PMNS}
}
where only two phases out of three phases $e^{i\delta_{a,b,c}}$ are independent. 
Out of three $e^{i\delta_{a,b,c}}$, we choose one freely to match to physics of the problem.  As mentioned before, Figs. \ref{fig:IntCP}\,(c) and (d) do not have the interference term with (a). 
So, we choose the Majorana phase of the dominant SM lepton $e^{i\delta_0}$ such that it does not have the phase dependence on  $e^{i\delta_0}$ in the interference with Fig. \ref{fig:IntCP}\,(a).
If we assume that the third generation Yukawa couplings dominate, $ f^{eff}_i = f^{eff}_3 \delta_{i3} $
we then obtain the expression
 \begin{equation}
\epsilon_L^{N_0}(W) =
 - \frac{\alpha_{\rm em}}{\sqrt{2}\sin^2\theta_W} \frac{\Delta m_h^2 }{m_0^2}
~{\rm Im} \left[ c_2c_3 e^{i \delta_c} + c_1s_2s_3e^{i (\dell + \delta_c)} \right]
\times\Big[1+{\rm ln}\left(\frac{m_W^2}{m_0^2}\right)  
\Big].
\label{eq:asymmExp3}
\end{equation} 
Here we see that the CP asymmetry is directly related to the PNMS phases.
The first factor of  Eq.~(\ref{eq:asymmExp2}) is about $10^{-3}$, since the asymmetry is enhanced by the smallness of the
$W$ mass. This we will see is an advantage and not a problem since sphaleron transitions are suppressed 
during the EW symmetry breaking epoch and we can realize baryogenesis even if only a small
fraction of the lepton number is converted into baryon number.

%%%%%%%%%%%%%%%%%%
\section{Relation of the phases}

Now we can relate the phases in our plan of spontaneous CP violation \cite{LeeTD73} with one complex VEV, \ie the phase of $\langle X\rangle$. Following the argument of Ref. \cite{KimEPJC15}, we can conclude that there will be no observable lepton asymmetry if $\delx=0$. Therefore, all the interference terms in Eq. (\ref{eq:asymmExp}) must have factors of the form $\sin(N_{ij}\delx)$ where $N_{ij}$ is an integer.
For example, consider the imaginary part of a specific term in  Eq. (\ref{eq:asymmExp}) before taking the sum with $i$ and $j$. From the product of (b) and (a)$^*$ of Fig. \ref{fig:IntCP}, we read one convenient term, \ie for $i=3$ and $j=1$, which has the overall phase $e^{i[\dell+\delta_a -n_1\delx+\delta_0]+i[n_3\delx-\delta_0]}$ 
where $\dell$ and $\delta_{a}$ are defined in Eq. (\ref{eq:PMNS}). The Majorana phase $\delta_0$ is the phase of the heavy lepton sector, which does not appear in this phase expression with  $i=3$ and $j=1$  if the lightest neutral heavy lepton dominates in the lepton asymmetry.  The imaginary part of this term is
\dis{
\sin[\dell+\delta_{a}-(n_1-n_3)\delx].\label{eq:PhaseRel}
}
In  Ref. \cite{KimEPJC15}, we argued that the observable phase $\dell$ in low energy experiments must be integer multiples of $\delx$ since there will be no electroweak scale CP violation effects if $\delx=0$ and $\pi$.  Along this line, we argue that $\dell=n_P\delx $ and $\delta_a=n_a\delx$, which are sufficient for the physical requirement.  In this case, Eq. (\ref{eq:PhaseRel}) becomes $\sin[(n_P+n_a-n_1+n_3)\delx]$.  Now, consider the sum with $i$ and $j$. We observe that each term has  the form of $Ae^{i(\pm n_P\delta_X+\delta')}+Be^{i\delta'}$ where $A$ and $B$ are real numbers formed with real angles and  $\delta'=n'\delx=n_a \delx,n_b \delx$, or $n_c \delx$, viz.  Eq. (\ref{eq:PMNS}). It is of the form
\dis{
&\left\{ A\cos[(\pm n_P+n')\delx]+B\cos[n'\delx]\right\} +i\left\{ A\sin[(\pm n_P+n')\delx]+B\sin[n'\delx]\right\}\\
&=\sqrt{\left\{ A\cos[(\pm n_P+n')\delx]+B\cos[n'\delx]\right\}^2+\left\{ A\sin[(\pm n_P+n')\delx]+B\sin[n'\delx]\right\}^2 }~e^{i\delta_{ij}} % \\ &
\equiv a_{ij}\,e^{i\delta_{ij}}\label{eq:PMNSeps}
   }  
which has the phase $\delta_{ij}=\arctan\left(\{ A\sin[(\pm n_P+n')\delx]+B\sin[n'\delx] \}/ \{ A\cos[(\pm n_P+n')\delx]+B\cos[n'\delx] \}
\right) $. Thus, every term has the vanishing phase if $\delx= 0$ and $\pi$. Thus, the sum in  Eq. (\ref{eq:PhaseRel}) gives 0 if $\delx= 0$ and $\pi$.  Even at this stage, we have obtained an important conclusion: the phases in the heavy lepton sector does not appear. For further relations, we must use a specific model relating $n_P, n',n_i$, and $n_j$, as we used the flipped-SU(5) model in relating $\dell$ and $\delq$ \cite{KimEPJC15}. Thus, the asymmetry takes a form,
\dis{
\eL^{N_0}(W)&
\approx \frac{\alpha_{\rm em}}{2\sqrt{2}\sin^2\theta_W}  \frac{\Delta m_h^2 }{m_0^2}
\sum_{i,j}{\cal A}_{ij}\sin[(\pm n_P+n'-n_i+n_j)\delx],
}
where ${\cal A}_{ij}$ are $a_{ij}$ times appropriate ratio of Yukawa couplings.  Note that there are only two independent $n'$ as commented before, below Eq. (\ref{eq:PMNS}).

 %%%%%%%%%%%%%%%%%
\section{Sphaleron processes during the EW broken phase}

Contrary to simple approximations, sphaleron transitions \cite{EWgen} are suppressed, but not vanishing when the
electroweak symmetry is broken. For a relatively large range of Higgs VEVs, as long as
$  v \leq T $, one can obtain at least a partial conversion of $L$ into $B$. 
Indeed the sphaleron rate in the equilibrium broken phase is given by \cite{Burnier:2005hp, D'Onofrio:2014kta}
\begin{equation}
 \Gamma_{\rm sph}^{\rm broken} = \kappa \alpha_W^4 T^4 \left(\frac{ 4\pi v}{g_W T} \right)^7 e^{-\frac{E_{\rm sph}}{T}}
 \end{equation}
 where $ \kappa $ is a constant, $ g_W, \alpha_W $ are the electroweak coupling and
 coupling strength and $ E_{\rm sph} $ the
 energy of the sphaleron energy barrier, proportional to the Higgs VEV, $ E_{\rm sph} = 1.52 4 \pi v/g_W $.
So, as found in  \cite{D'Onofrio:2014kta}, in the SM with a Higgs mass of 125 GeV, 
the sphaleron processes remain in thermal equilibrium until one reaches temperatures of the order 
$ T_* = (131.7 \pm 2.3)$ GeV, where $ \frac{v}{T} > 1 $.
In case the Higgs VEVs remain non-vanishing, as we advocate here, such VEVs are proportional to 
the temperature in the high $T$ regime, $ v(T) = \sqrt{ v(0)^2 + k^2 T^2} $, as discussed in \cite{Senj79}. 
Therefore, for $ k \sim 1 $ the sphaleron processes may enter equilibrium for low temperatures 
above the electroweak scale $ T \geq v(0) $ as long as
\begin{equation}
\frac{\Gamma_{\rm sph}^{\rm broken}}{T^3 H(T)} = 
\kappa \alpha_W^4 \left(\frac{ 4\pi k}{g_W} \right)^7 e^{- 1.52 k \frac{4\pi}{g_W}} \sqrt{\frac{90}{\pi^2 g_*}} 
\frac{M_P}{T} \geq 1 \; .
 \end{equation}

In case the sphaleron processes are active only for a very short range of temperatures, nevertheless a partial conversion of lepton number into baryon number may still be possible, giving rise to the observed baryon asymmetry if the lepton number is sufficiently large.
 
 %%%%%%%%%%%%%%%%%%%%
 \section{Conclusion}
By introducing only one CP phase by a complex VEV  of a SM singlet field $X$ and assuming leptogenesis via the lightest Majorana neutrino out of equilibrium decay 
during a phase where the electroweak symmetry is broken, we  have that the dominant contribution
to the CP asymmetry in the early Universe arises from a $W$-boson loop, directly containing
the PMNS phase  $\dell$.
In this way we are able to have a novel mechanism to relate  high and low energy 
CP violation, in particular in the case when a single CP phase is introduced by spontaneous 
mechanism at a high energy scale along the Froggatt-Nielsen method.

Even in case more phases are present and a non-vanishing contribution from the classical
loop with the heavier RH neutrinos states is present, the PMNS phase could still represent
the dominant part and allow for a direct correlation of the baryon asymmetry to neutrino
observables.

 %%%%%%%%%%%%%%%%%%%% 
\acknowledgments{We thank S.M. Barr for useful comments. 
L.C. acknowledges partial financial support by  the European Union's Horizon 2020 research and 
innovation programme under the Marie Sklodowska-Curie grant agreements  No 690575 and No 674896, 
J.E.K. is supported in part by the National Research Foundation (NRF) grant funded by the Korean Government (MEST) (NRF-2015R1D1A1A01058449) and  the IBS (IBS-R017-D1-2016-a00), and B.K.  is supported in part by the NRF-2013R1A1A2006904 and Korea Institute for Advanced Study grant funded by the Korean government.}

\end{document}